\documentclass[aps, prl, preprint, superscriptaddress, nobibnotes, amsmath, amssymb, tightenlines
]{revtex4-2}

\usepackage{hyperref}
    \hypersetup{
        colorlinks=true,
        linkcolor=blue,
        urlcolor=blue,
        citecolor=blue,
        bookmarksopen=true,
        bookmarksnumbered,
    }
\usepackage[all]{hypcap}
\usepackage[normalem]{ulem}
\usepackage{silence}
\usepackage{graphicx}
\usepackage{latexsym}
\usepackage{bm}
\usepackage{siunitx}

\usepackage{zref-xr, zref-user}
\zexternaldocument*{TaSe3_SI}

\usepackage{etoolbox}
\usepackage{bookmark}
\makeatletter
\newcommand{\listoffiguresbookmarks}{%
  \bookmarksetup{level=0}
  \@starttoc{lofb}}
\makeatother
\makeatletter
\pretocmd\endfigure{%
\addtocontents{lofb}{%
  \protect{%
    \bookmark[
    rellevel=0,
    keeplevel,
    dest=\@currentHref,
    ]{Fig. \thefigure: \@currentlabelname}}}%
}{}{\errmessage{Patching \noexpand\endfigure failed}}
\makeatother

\newcommand*{\refer}[2]{\hyperref[#1]{\ref{#1}#2}}
\newcommand*{\wave}{$\sim$}

\newcommand*{\EF}{$E_\text{F}$}
\newcommand*{\TS}{$\text{TaSe}_3$}
\newcommand*{\kk}[1]{$k_\text{#1}$}
\newcommand*{\Plane}{$(\bar{1}01)$}
\newcommand*{\Gama}{$\widetilde{\Gamma}$}
\newcommand*{\X}{$\widetilde{\text{X}}$}
\newcommand*{\A}{$\si{\angstrom^{\text{-}1}}$}
\newcommand*{\TSS}[1]{TSS$_\text{#1}$}
\newcommand*{\IS}{$in$-$situ$}

\usepackage{xcolor}
\definecolor{green}{rgb}{0, 0.7, 0}

\begin{document}

\title{Visualization of the strain-induced topological phase transition \texorpdfstring{\\}{}in a quasi-one-dimensional superconductor TaSe\texorpdfstring{$_3$}{3}}

\author{Chun~Lin}
    \affiliation{Institute for Solid State Physics, University of Tokyo, Kashiwa, Chiba 277-8581, Japan.}
\author{Masayuki~Ochi}
    \affiliation{Department of Physics, Osaka University, Toyonaka,~Osaka 560-0043, Japan.}
\author{Ryo~Noguchi}
\author{Kenta~Kuroda}
    \affiliation{Institute for Solid State Physics, University of Tokyo, Kashiwa, Chiba 277-8581, Japan.}
\author{Masahito~Sakoda}
    \affiliation{Department of Applied Physics, Hokkaido University, Kita-ku, Sapporo 060-8628, Japan.}
\author{Atsushi~Nomura}
    \affiliation{Department of Physics, Tokyo University of Science, Shinjuku-ku, Tokyo 162-8601, Japan.}
\author{Masakatsu~Tsubota}
    \affiliation{Department of Physics, Gakushuin University, Toshima-ku, Tokyo 171-8588, Japan.}
\author{Peng~Zhang}
\author{Cedric~Bareille}
\author{Kifu~Kurokawa}
\author{Yosuke~Arai}
\author{Kaishu~Kawaguchi}
\author{Hiroaki~Tanaka}
 \affiliation{Institute for Solid State Physics, University of Tokyo, Kashiwa, Chiba 277-8581, Japan.}
\author{Koichiro~Yaji}
    \affiliation{Institute for Solid State Physics, University of Tokyo, Kashiwa, Chiba 277-8581, Japan.}
    \affiliation{Research Center for Advanced Measurement and Characterization, National Institute for Materials Science, Ibaraki 305-0003, Japan.}
\author{Ayumi~Harasawa}
    \affiliation{Institute for Solid State Physics, University of Tokyo, Kashiwa, Chiba 277-8581, Japan.}
\author{Makoto~Hashimoto}
\author{Donghui~Lu}
    \affiliation{Stanford Synchrotron Radiation Lightsource, SLAC National Accelerator Laboratory, Menlo Park, California 94025, USA.}
\author{Shik~Shin}
    \affiliation{Office of University Professor, University of Tokyo, Kashiwa, Chiba 277-8581, Japan.}
\author{Ryotaro~Arita}
    \affiliation{RIKEN Center for Emergent Matter Science (CEMS), Wako, Saitama 351-0198, Japan.}
    \affiliation{Department of Applied Physics, University of Tokyo, Tokyo 113-8656, Japan.}
\author{Satoshi~Tanda}
    \affiliation{Department of Applied Physics, Hokkaido University, Kita-ku, Sapporo 060-8628, Japan.}
    \affiliation{Center of Education and Research for Topological Science and Technology, Hokkaido University, Kita-ku, Sapporo 060-8628, Japan.}
\author{Takeshi~Kondo}
\email[]{kondo1215@issp.u-tokyo.ac.jp}
    \affiliation{Institute for Solid State Physics, University of Tokyo, Kashiwa, Chiba 277-8581, Japan.}
    \affiliation{Trans-scale Quantum Science Institute, The University of Tokyo, Bunkyo, Tokyo 113-0033, Japan}
    
\maketitle

\pdfbookmark[1]{Introduction}{Introduction}

\textbf{
Control of the phase transition from topological to normal insulators can allow for an on/off switching of spin current. 
While topological phase transitions have been realized by elemental substitution in semiconducting alloys, such an approach requires preparation of materials with various compositions, thus it is quite far from a feasible device application, which demands a reversible operation. 
Here we use angle-resolved photoemission spectroscopy (ARPES) and spin-resolved ARPES to visualize the strain-driven band structure evolution of the quasi-1D superconductor \TS{}. 
We demonstrate that it undergoes reversible strain-induced topological phase transitions from a strong topological insulator phase with spin-polarized, quasi-1D topological surface states, to topologically trivial semimetal and band insulating phases. 
The quasi-1D superconductor \TS{} provides a suitable platform for engineering the topological spintronics, for example as an on/off switch for spin current robust against impurity scattering.
}

Since the discovery of topological insulators (TIs)~\cite{Fu_Phys.Rev.Lett.9810_2007,Fu_Phys.Rev.B764_2007,Hasan_Rev.Mod.Phys.824_2010,Qi_Rev.Mod.Phys.834_2011,Ando_J.Phys.Soc.Jpn.8210_2013}, controlling the fascinating properties has been attempted. 
The most direct way for it is to utilize the phase transition from the topological to normal insulators, which has been so far realized by elemental substitution in semiconducting alloys \cite{Xu_Science3326029_2011,Sato_Nat.Phys.711_2011,Xu_Nat.Commun.31_2012,Brahlek_Phys.Rev.Lett.10918_2012,Zhang_Science3396127_2013,Wu_Nat.Phys.97_2013,Xu_Nat.Commun.61_2015}; the spin-orbit coupling (SOC) and lattice constant are both varied simultaneously, leading to the band inversion or eliminating it. 
To enable a reversible control of the phase transition required for a device application, utilizing the strain effect, which can tune the lattice constant,  
would be a better and simpler approach; for that, we need to search for a suitable material, which has the topological properties and can easily  change its lattice constant by strain.

In the transition-metal trichalcogenides $MX_3$  (\textit{M} = Nb, Ta; \textit{X} = S, Se), physical properties vary and the electronic orders evolve differently in accordance with the distinctive stacking sequences of the 1D chain variants~\cite{Monceau_Adv.Phys.614_2012}.
Among the series of $MX_3$, \TS{} is especially appealing in that superconductivity emerges at low temperatures ($\sim 2$ K)~\cite{Sambongi_J.Phys.Soc.Jpn.424_1977}, differently from other members which typically undergo the charge density wave (CDW) transitions~\cite{Monceau_Adv.Phys.614_2012}. Owing to the quasi-1D metallic character, it is proposed from the application point of view that \TS{} is suitable for the downscaled local interconnectors in electronic devices~\cite{Stolyarov_Nanoscale834_2016,Liu_NanoLett.171_2017}. More intriguingly, recent $ab$ $initio$ calculations predict that semimetallic \TS{} belongs to a $Z_\text{2}$ strong TI phase, and it has been raised as a candidate of topological superconductor~\cite{Nie_Phys.Rev.B9812_2018}  advantageous over other compounds~\cite{Fu_Phys.Rev.Lett.1059_2010,Sasaki_Phys.Rev.Lett.10721_2011,Wang_Science3366077_2012,Mourik_Science3366084_2012,Sakano_Nat.Commun.61_2015,Guan_Sci.Adv.211_2016,Zhang_Science3606385_2018,Clark_Phys.Rev.Lett.12015_2018,Liu_Nat.Commun.111_2020} in that the TSSs are formed at the Fermi level (\EF{}), the superconductivity occurs in stoichiometric crystals without suffering from the doping-induced inhomogeneity, and the crystal structure is built from the van der Waals stacking suitable for the application~\cite{Yuan_Nat.Phys.1510_2019}. Furthermore, this compound is in proximity to other topological phases, which thus potentially bring an attractive functionality of controlling the fascinating topological properties by the fine-tuning of a single physical parameter~\cite{Nie_Phys.Rev.B9812_2018}.

The electronic structure of \TS{} has been shown by ARPES more than one decade ago~\cite{Perucchi_Eur.Phys.J.B394_2004} and in a recent study with much better resolutions~\cite{Chen_Matter36_2020}. We use various types of ARPES techniques to observe the band structure of \TS{}. 
In particular, the spin-resolved ARPES (SARPES) with high momentum and energy resolutions is indispensable to identify the topological nature of this compound. Hence, we employ a laser-based SARPES which satisfies such requirements, and  unambiguously reveal for the first time that \TS{} is indeed in a strong TI phase as theory predicts. Moreover, two-step phase transitions are demonstrated by means of an \IS{} strain control of the band structure:  with applying tensile strain to the samples, we observe a drastic evolution of band structure demonstrating the topological phase transition from a strong TI phase to a trivial semimetal phase; upon further increasing of the strain, the system eventually becomes a trivial band insulator, realizing the metal-insulator transition. Here we emphasize that the transition from topologically non-trivial to trivial state is unique in TaSe$_3$; while the strain control of the topological state was previously attempted for Bi$_2$Se$_3$~\cite{Flototto_NanoLett.189_2018}, the variation of the band structure by strain was rather small, exhibiting only a slight energy shift of Dirac point. The uniaxial strain method was also effectively used to align the nematic superconducting domains in Sr-doped Bi$_2$Se$_3$~\cite{Kostylev_Nat.Commun.111_2020}; such an effect, however, is not drastic in energy scale as that in the topological phase transition as realized in TaSe$_3$. ZrTe$_5$ is another good system to control the topological phase by strain~\cite{Mutch_Sci.Adv.58_2019,Zhang_Nat.Commun.121_2021}, whereas the transition proposed was between two non-trivial phases (weak TI and strong TI phases), differing from the present results. The reversible control from the topological to non-topological phases in a single compound TaSe$_3$ without the necessity of elemental substitution in alloys opens up a new path toward applications in spintronics or optoelectronics. In particular, the controlling of topological phase transition accompanied by superconductivity will bring whole new ideas for future applications.

\pdfbookmark[1]{Crystal and synchrotron-ARPES}{Crystal and synchrotron-ARPES}

\begin{figure}[!tb]
    \includegraphics[width=\textwidth]{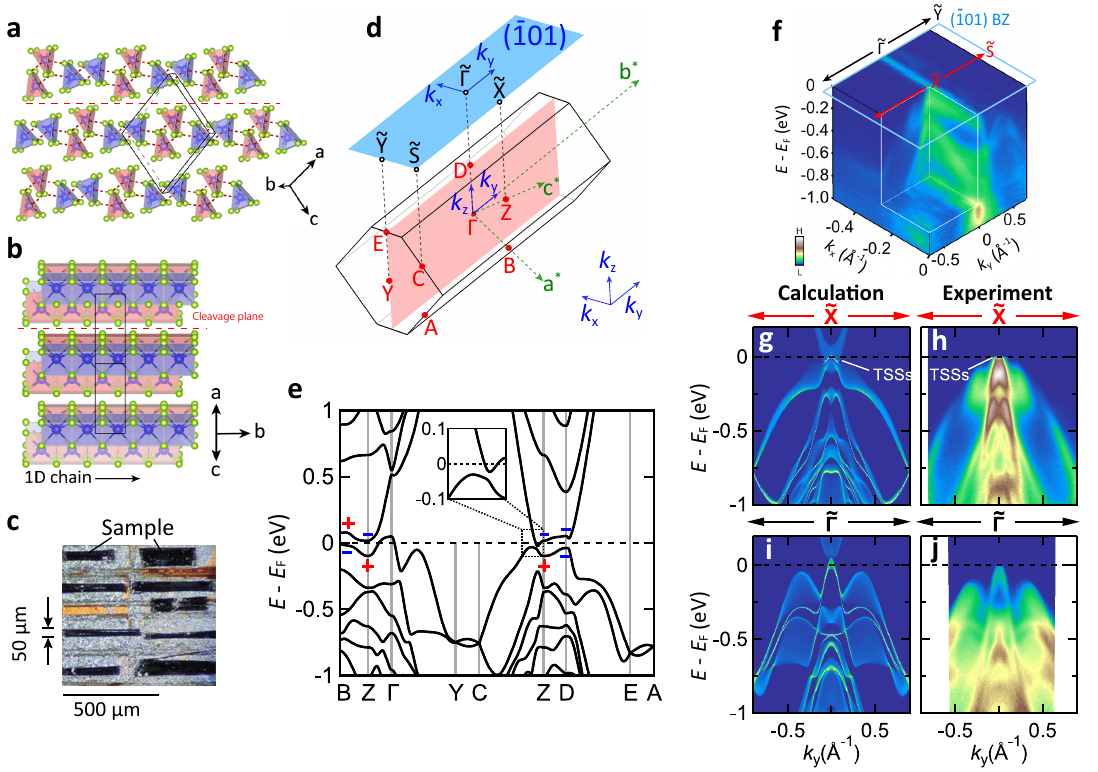}
    \caption[Crystal and electronic structure of TaSe3]{
    Crystal structure of \TS{} and its electronic structure revealed by ARPES.
    \textbf{a}, Crystal structure of \TS{}, stacked by two types of 1D chains with different bond lengths, as indicated by the reddish and bluish prisms, respectively.
    Adjacent 1D chains have an offset of b/2 along chain direction and the bondings between them are shown by the dark red dashed lines forming the \Plane{} layers stacked by the van der Waals force.
    \textbf{b}, The side view of the crystal structure along the chain.
    The unit cell is indicated by the black lines and the light red dashed lines mark the \Plane{} natural cleavage plane.
    \textbf{c}, The image of the samples used for the ARPES measurements indicating the typical crystal size with \wave{} 500 \si{\um} in length and 50 - 100 \si{\um} in width.
    \textbf{d}, 3D bulk Brillouin zone (BZ) and the projected \Plane{} surface BZ.
    \kk{x} and \kk{y} are in-plane with the former perpendicular to the chain direction and the latter parallel to it, \kk{z} is along out-of-plane direction.
    \textbf{e}, Electronic structure of \TS{} with spin-orbit coupling.
    The parity eigenvalues at the B, Z, and D points are indicated explicitly.
    \textbf{f}, 3D plot of the band dispersion on the \Plane{} surface BZ probed by 50 eV photons.
    \textbf{g},\textbf{h}, Calculated topological surface states (TSSs) and experimental band dispersion at the \X{} point, respectively.
    The possible TSSs in \textbf{h} are also indicated.
    \textbf{i},\textbf{j}, Calculated surface states (SSs) and experimental band dispersion at the \Gama{} point, respectively.
    The photon energies used in \textbf{h} (50 eV) and \textbf{j} (30 eV) were selected so that the SSs were better resolved.
    See Supplementary Notes~\zref{hvGamma},\zref{hvX} and Figs.~\zref{hvGammaFig},\zref{hvXFig} for more detailed discussions about the spectra obtained using different photon energies and polarizations. 
    \label{ARPES}
    }
\end{figure}

The \TS{} crystal is built from the stacking of layers bonded together by the weak van der Waals force (Figs.~\refer{ARPES}{a,b});  this expects the \Plane{} plane to be left on the surface after the sample cleaving (Fig.~\refer{ARPES}{c}), which has indeed been confirmed by our X-ray diffraction measurements (see Supplementary Note~\zref{Samples} and Fig.~\zref{XRDFig}).  The band structure calculated along high-symmetry momentum lines (see Fig.~\refer{ARPES}{d}) is plotted in Fig.~\refer{ARPES}{e}; 
band inversion occurs around the Z point, opening an energy gap induced by the SOC effect (the inset of Fig.~\refer{ARPES}{e}).
We have used synchrotron-ARPES with tunable photon energies to experimentally clarify the overall electronic structure of \TS{}. 
Figure~\refer{ARPES}{f} exhibits the overview of band dispersions observed on the \Plane{} surface Brillouin zone (BZ; see Fig.~\refer{ARPES}{d}).  In Figs.~\refer{ARPES}{g,h} and Figs.~\refer{ARPES}{i,j}, we compare 
the calculated surface-state spectra (left panels) and ARPES dispersions (right panels) across the \X{} and \Gama{} points, respectively; a good agreement between the two is confirmed regardless of the momentum cuts. 
 
The surface and bulk states have distinct spectral properties and can be selectively enhanced 
by using proper photon energies and light polarizations (see Supplementary Notes~\zref{hvGamma},\zref{hvX} and Figs.~\zref{hvGammaFig},\zref{hvXFig}). 
In Supplementary Figs.~\zref{hvGammaFig}m,n, we plot the Fermi surface mappings and the constant-energy mappings at a high binding energy ($E_{\rm B}$=0.6 eV), respectively, with different \kk{z}'s accessed by changing the photon energies.
Unlike the mapping at $E_{\rm B}$=0.6 eV, the Fermi surface is found to be non-dispersive along \kk{z}, suggesting that the low-energy sharp features seen in Fig.~\refer{ARPES}{h} very likely come from surface states.
The TSSs are expected, by calculations, to be confined in a limited momentum area around the \X{} point and within a small bulk gap opened at \EF{}; thus, ultra-high energy and momentum resolutions are required in ARPES experiments to visualize the predicted TSSs, and most importantly, the spin-resolved measurements in high-resolutions are indispensable to identify that the surface states observed have indeed a topological origin.

\pdfbookmark[1]{TSSs and laser-ARPES}{TSSs and laser-ARPES}

\begin{figure}[!]
    \includegraphics[width=\textwidth]{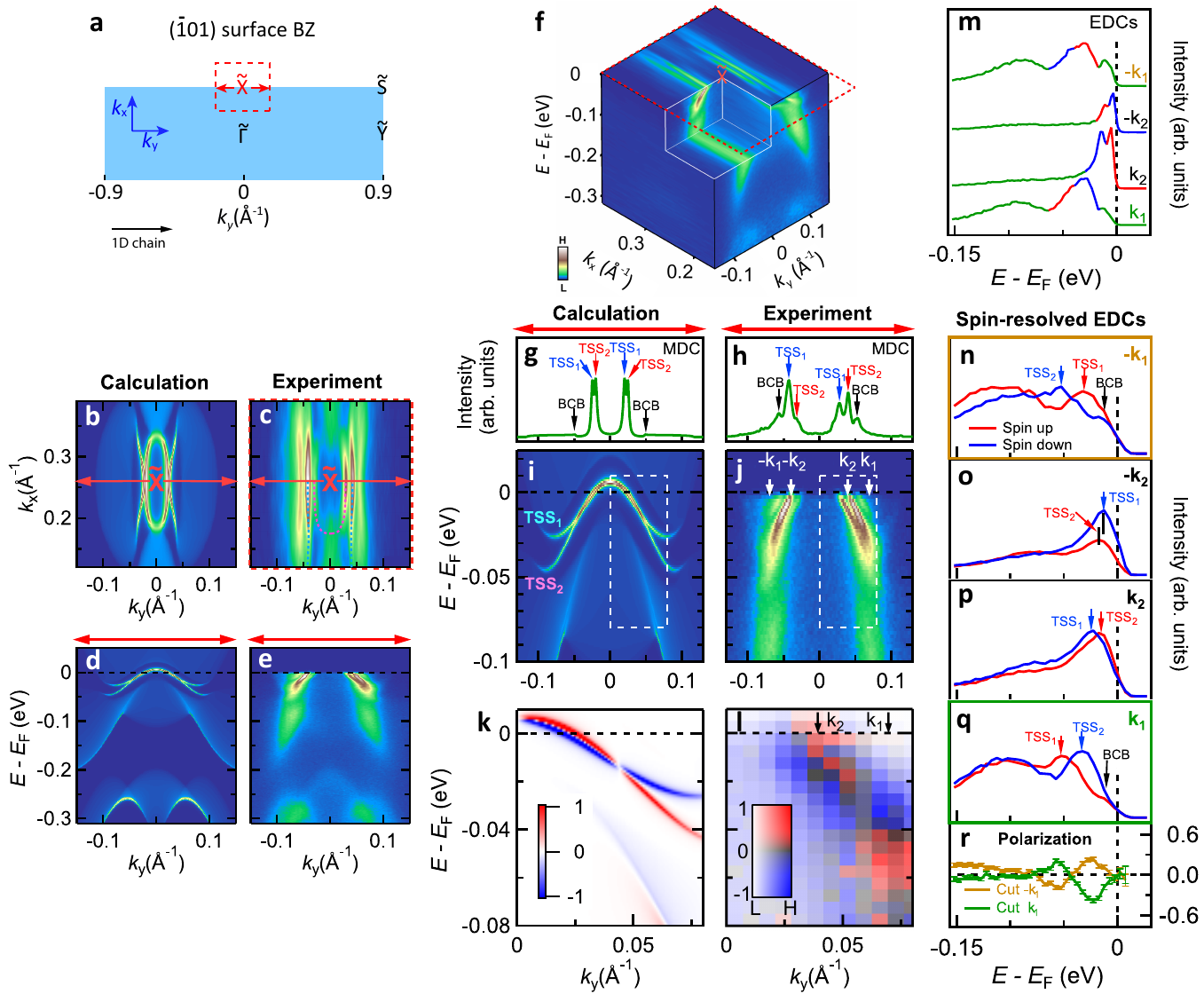}
    \caption[TSSs near the X point in TaSe3 revealed by the laser-SARPES with 7 eV photons]{
    TSSs near the \X{} point on the \Plane{} surface BZ revealed by the laser-SARPES with 7 eV photons.
    \textbf{a}, Schematic of the \Plane{} surface BZ.
    \textbf{b},\textbf{c}, Calculated and experimental surface-state Fermi surfaces within the momentum range around the \X{} point specified by the red dashed rectangle in \textbf{a}.
    The dashed curves in \textbf{c} depict the contour of the TSSs.
    \textbf{d},\textbf{e}, Calculated and experimental surface-state dispersions at the \X{} point with momentum positions indicated by the red arrows in \textbf{b},\textbf{c}.
    \textbf{f}, 3D plot of the experimental surface-state dispersion around the \X{} point.
    The spectra in \textbf{c},\textbf{f} are symmetrized with respect to the zone edge.
    \textbf{g},\textbf{h}, Calculated and experimental momentum distribution curves (MDCs) at \EF{} along the red arrows in \textbf{b,c}, respectively, indicating the correspondences of the two branches of the TSSs (\TSS{1}, \TSS{2}) and bulk conduction bands (BCBs).
    \textbf{i},\textbf{j}, Magnified calculated and experimental TSSs for \textbf{d,e}, respectively.
    \textbf{k},\textbf{l}, Calculated spin texture and experimental spin-polarization map, respectively, with energy-momentum range specified by the white dashed rectangles in \textbf{i},\textbf{j}.
    The spin-polarization map in \textbf{l} is coded by a 2D color scale with horizontal and vertical axes indicating the spectral weight and the polarization, respectively.
    The spin-polarization vector is nearly in-plane and perpendicular to the chain direction; see complete spin-polarization components in Supplementary Fig.~\zref{SpinComponentFig}.
    \textbf{m-q}, Spin-integrated (\textbf{m}) and spin-resolved (\textbf{n-q}) energy distribution curves (EDCs) of the TSSs at four momentum cuts indicated in \textbf{j}.
    \textbf{r}, Representative spin polarization magnitude for outer two cuts -k$_1$ and k$_1$.
    \label{Laser}
    }
\end{figure}

The fine details of the low-energy electronic structure near \X{} on the \Plane{} surface BZ has been examined at high-resolutions by a 7-eV laser-ARPES equipped with a 3D spin-resolved detector. Now we focus on the momentum area of a red dashed square in Fig.~\refer{Laser}{a} surrounding the \X{} point; calculations predict that the TSSs, characterized by sharp spectra near \EF{}, emerge between the bulk conduction and valence bands (BCB and BVB) when approaching the \X{} point, where these bands are inverted.  
In Figs.~\refer{Laser}{b,c} and Figs.~\refer{Laser}{d,e}, we compare the Fermi surfaces and dispersions across the \X{} point, respectively, obtained by calculations (left panels) and ARPES (right panels); the ARPES data are also three-dimensionally displayed in Fig.~\refer{Laser}{f} for better understanding.
The low-energy spectra measured by laser-ARPES (Figs.~\refer{Laser}{c,e,f}) clearly disentangle fine structures near the \X{} point, which agree well with the calculated TSSs in the strong TI phase (Figs.~\refer{Laser}{b,d}). 
To examine more details of the TSSs, we magnify the calculated and ARPES dispersions across \X{} in Figs.~\refer{Laser}{i,j}, respectively. The momentum distribution curves (MDCs) extracted at \EF{} (Figs.~\refer{Laser}{g,h}) identify two pairs of TSSs (TSS$_1$ and TSS$_2$) and one pair of BCBs, indicating a good consistency between calculations and ARPES measurements.
As detailed in Supplementary Fig.~\zref{TSSBCBEDCsFig}, the band dispersion of TSS$_1$ goes up in energy with increasing $k$ and merges into the BCB around $k = 0.1$ \A{}, whereas that of TSS$_2$ goes down and merges into the BVB, just as expected for a strong TI phase (Fig.~\refer{Laser}{i}).
We also note here that spectra assigned to BCBs have relatively sharp peaks, unlike the continuum intensities for the surface band calculations (Fig.~\refer{Laser}{i}), because of the following reason: ARPES at one photon energy, in principle, captures the band along a $k_x$-$k_y$ sheet at a specific $k_z$, thus it can exhibit sharp peaks in the spectra not only for the surface band but also for the bulk band, while the latter could be suppressed by the $k_z$ broadening in photoemission. This contrasts with the surface calculations, which project the bulk band spectra over the whole range of $k_z$.

To finalize our conclusion that \TS{} belongs to a strong TI phase, we demonstrate here that the surface states we observed are spin-polarized just as expected by calculations. For this purpose, the experimental setting is now switched to the spin-resolved mode. Figure~\refer{Laser}{l} plots the SARPES data which map the spin $k_x$ component (slightly off from the $k_x$ direction in reality due to experimental geometry), measured for an energy-momentum region surrounded by a white box in Fig.~\refer{Laser}{j}; one will see how high the momentum resolution is in laser-SARPES compared with that in synchrotron-ARPES (Fig.~\refer{ARPES}{h}).
To examine more details, we extract, in Figs.~\refer{Laser}{n\textendash q}, the spin-resolved energy distribution curves (EDCs) at the four specific $k$ points indicated by arrows in Fig.~\refer{Laser}{j}; the observed surface states are clearly spin-polarized (see Supplementary Fig.~\zref{SpinComponentFig} for the complete spin-polarization components).
In Fig.~\refer{Laser}{r}, the magnitude of spin-polarization is quantified for two outer $k$ points with opposite signs (+$k_1$ and -$k_1$). The spin-polarization is reversed both with different energies and between $\pm k$. Intriguingly, we also found that these relationships on reversal are also swapped between different $|k|$ ($\pm k_1$ and $\pm k_2$), which yields a nodal point with spin degeneracy possibly due to the band hybridization of TSSs. 
Note here that, while the spectral intensities in Fig.~\refer{Laser}{o} are dominated by the down spin (blue spectrum) mostly due to the matrix element effect, the spin degeneracy lifted by \wave{} 10 meV is clearly resolved, validating the spin-texture reversal. For clarity, the conclusion experimentally obtained is summarized in Fig.~\refer{Laser}{m} by painting the spin-integrated EDCs with red and blue colors assigned for up- and down-spin, respectively.  
All these features in the spin texture are well reproduced by calculations (Fig.~\refer{Laser}{k}), justifying that our experiments successfully captured the spin-momentum locked TSSs, and hence established a strong TI phase in \TS{}. We note that \TS{} is not an insulator with a fully opened bulk gap in its strong TI phase, but a semimetal with a finite density of states at the Fermi level, which is a required condition to realize superconductivity making this compound all the more fascinating for the possible applications with topological superconductivity.

\pdfbookmark[1]{Strain-induced phase transitions}{Strain-induced phase transitions}

\begin{figure}[!]
    \includegraphics[width=\textwidth]{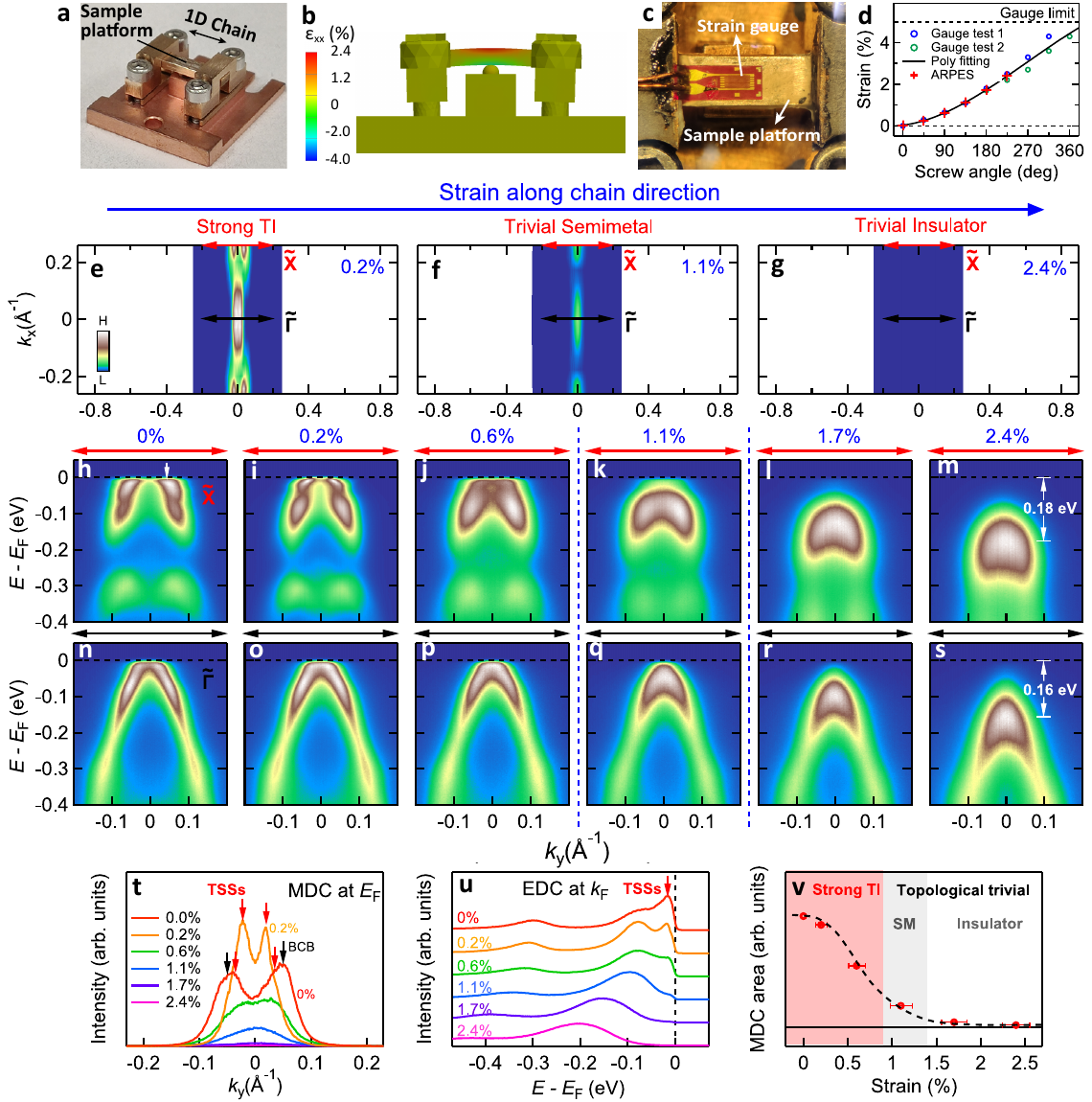}
    \caption[Topological phase transition driven by the in situ controlled tensile strain along chain]{
    Observed topological and metal-insulator phase transitions driven by the \IS{} controlled tensile strain along the chain.
    \textbf{a}, The strain device.
    The samples were mounted on the sample platform with 1D chain along the length.
    \textbf{b}, Side view of the device with color encoding simulated longitudinal strain distribution ($\epsilon_\text{xx}$).
    \textbf{c}, The commercial strain gauge attached to a sample platform.
    \textbf{d}, Strain magnitude versus screw angles for two representative tests measured by the strain gauges.
    The red markers indicate the strain magnitude in the ARPES measurements obtained by interpolating the fitting curve at the corresponding screw angles controlled \IS{}.
    \textbf{e-g}, Measured Fermi surfaces in the strong TI, trivial semimetal (SM), and trivial insulator phases with strain of 0.2\%, 1.1\%, and 2.4\%, respectively.
    The spectra are symmetrized with respect to \kk{x} = 0.
    \textbf{h-m}, Dispersions at the \X{} point with increasing strain indicated on the top of each panel.
    \textbf{n-s}, Corresponding dispersions at the \Gama{} point.
    The energy positions of the intensity maxima are indicated in \textbf{m,s}.
    \textbf{t},\textbf{u}, MDCs at \EF{} (\textbf{t}) and EDCs at \kk{F} (\textbf{u}) near the \X{} point with different strain.
    The \kk{F} position of the TSSs is indicated by the white arrow in \textbf{h}.
    \textbf{v}, Evolution of the spectral intensity at \EF{} (MDC area in \textbf{t}) with different strain.
    The dashed line is the guide for the eyes.
    See Methods for more details.
    \label{Strain}
    }
\end{figure}

The quasi-1D structure is advantageous to modify the band structure effectively by uniaxial pressure. Hence, \TS{} built from chains is the ideal material to realize the strain-induced topological phase transition. Here we use the simplest technique of applying pressure to substances, which just mechanically bends a crystal on a substrate (see Figs.~\refer{Strain}{a,b})~\cite{Ricco_Nat.Commun.91_2018}, and reveal a systematic variation of the band structure under the tensile strain along the chain direction which is controlled \IS{}. 
The samples were mounted on the sample platform of the strain device shown in Fig.~\refer{Strain}{a}, by which tensile strain is applied via tightening four screws (Fig.~\refer{Strain}{b}). 
We measured the amount of strain at the sample position by commercial strain gauges (Fig.~\refer{Strain}{c}), and made a diagram relating the turning angle of screws and the resulting strain (Fig.~\refer{Strain}{d}), which was then used for the \IS{}  strain control. 
The reliability of this diagram was evaluated by simulating the strain distribution in the device with finite element analysis (Fig.~\refer{Strain}{b}); a good agreement has been confirmed between the simulation and the strain gauge measurements (see Methods for more information about the strain device and the estimation of the strain).

\begin{figure}[!tb]
    \includegraphics[width=\textwidth]{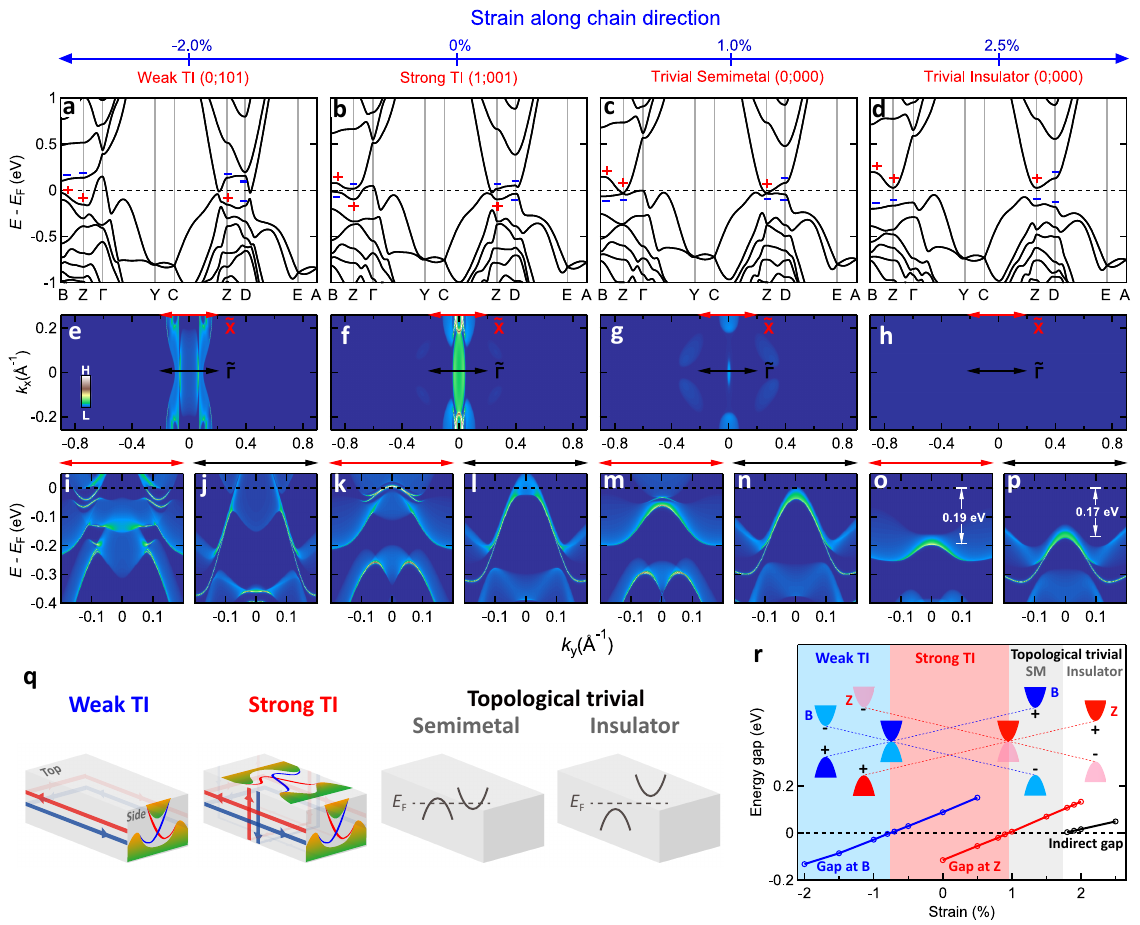}
    \caption[Calculated strain-induced topological phase transition]{
    Calculated topological phase transition and metal-insulator transition driven by the strain along chain direction.
    \textbf{a-d}, Electronic structures in the weak TI, strong TI, trivial semimetal, and trivial insulator phases under the strain values indicated on the top. 
    The parity eigenvalues at the B, Z, and D points where band inversions occur are indicated explicitly.
    \textbf{e-h}, Corresponding Fermi surfaces on the \Plane{} surface.
    \textbf{i-p}, Corresponding band dispersions at the \X{} (\textbf{i},\textbf{k},\textbf{m},\textbf{o}) and \Gama{} (\textbf{j},\textbf{l},\textbf{n},\textbf{p}) points with momentum ranges indicated by the red and black arrows in \textbf{e-h}, respectively.
    The energy positions of the intensity maxima are indicated in \textbf{o,p}, which nicely match the energy scales in Figs.~\refer{Strain}{m,s}.
    \textbf{q}, Schematics of the surface spin current, TSSs, or bulk band structure in different phases.
    The dispersions of TSSs are shown on the crystal surfaces (on all surfaces in the strong TI phase and only on the side surface in the weak TI phase).
    \textbf{r}, Calculated strain phase diagram of \TS{}.
    The direct gaps at the Z and B points, as well as the indirect gap (defined by the energy difference between conduction band bottom and valence band top) are plotted as a function of strain along the chain.
    The zero-energy crossing points of these energy scales define the quantum critical points (QCPs) of \wave{} -0.75\%, 0.95\%, and 1.75\% for phase transitions among weak TI, strong TI, trivial semimetal, and trivial band insulator phases.
    The schematics indicate the evolutions of bulk band energies and the parity inversions at the Z and B points across the topological QCPs.
    \label{StrainCalc}
    }
\end{figure}

Figures~\refer{Strain}{e\textendash g} present the evolution of Fermi surface mappings by laser-ARPES with increasing tensile strain along the chain (0.2 $\%$, 1.1 $\%$, and 2.4 $\%$, respectively), signifying two-step phase transitions:  the high intensities seen at low strain (Fig.~\refer{Strain}{e}) are significantly suppressed by applying higher strain (Fig.~\refer{Strain}{f}), indicating that the TSSs disappear and thus the system has transitioned from a strong TI phase to a trivial semimetal phase; its semimetallic nature will be confirmed by comparisons with calculations in Fig.~\ref{StrainCalc} and Supplementary Fig.~\zref{SemimetalFig}. The vague intensities left for the bulk states (Fig.~\refer{Strain}{f}) eventually vanish with further increase of the strain (Fig.~\refer{Strain}{g}), revealing a transition of the system to a trivial band insulator phase, which opens a bulk gap at \EF{}. 

More details of these two-step phase transitions are examined in Figs.~\refer{Strain}{h\textendash m} and Figs.~\refer{Strain}{n\textendash s}, which exhibit the evolution of band dispersions across \X{} and \Gama{} (red and black arrows in Figs.~\refer{Strain}{e\textendash g}), respectively, with gradually increasing tensile strain along the chain by the \IS{} control. 
The highly intense signals due to TSSs are distinguished close to \EF{} at zero strain (Fig.~\refer{Strain}{h}) and agree to the results in Fig.~\refer{Laser}{e}; the difference in the spectral sharpness between these data mostly comes from the different quality of cleaved surfaces and different light polarizations used in two different ARPES apparatuses each for the spin-texture and strain-control measurements (see Supplementary Figs.~\zref{TSSs_XFig},\zref{TSSsAllFig} for TSSs obtained at different conditions).
By applying the tensile strain, the hole bands around \X{} and \Gama{} both gradually shifted toward higher binding energies, demonstrating successful band-engineering by strain. The key features associated with the phase transitions are seen especially in the close vicinity of \EF{} around the \X{} point: the high intensities of TSSs seen at zero strain (Fig.~\refer{Strain}{h}) were substantially suppressed at around 1 $\%$ strain (Figs.~\refer{Strain}{j,k}), which was then followed by a gap opening with larger strain (Figs.~\refer{Strain}{l,m}). 
These behaviors are more directly demonstrated in Fig.~\refer{Strain}{t} and Fig.~\refer{Strain}{u} by extracting  MDCs at \EF{} across \X{} and EDCs at \kk{F} around \X{}, respectively, which reveal the spectral evolution of TSSs (marked by red arrows in these panels). As the strain was increased in the strong TI phase, the TSSs became confined into a smaller momentum area (Fig.~\refer{Strain}{t}), as a result of the reduced band inversion. Further increase of strain caused the spectral peaks of TSSs to disappear by eliminating the band inversion, and consequently the system transitioned to a trivial semimetal phase. Eventually, the strain got large enough to reopen the band gap with totally vanishing the spectral intensity around \EF{}, and thus the system transitioned to a trivial insulator phase. A previous study has shown an anomalous strain effect on the resistivity, which increased by several orders of magnitude when the applied strain along the chain reached \wave{} 0.6\%~\cite{Tritt_Phys.Rev.B3410_1986}, in agreement with our observation of the band structure and the depletion of the spectral intensity at \EF{} shown in Fig.~\refer{Strain}{v}.

\pdfbookmark[1]{Calculations and discussions}{Calculations and discussions}

To fully understand the mechanism of the phase transitions we observed, $ab$ $initio$ calculations were performed for \TS{} under the strain along the chain direction. The evolution of the bulk electronic structure in the calculations (Figs.~\refer{StrainCalc}{b-d}) nicely illustrates the mechanism of the phase transition induced by reducing and eliminating the band inversion with increasing tensile strain, which are manifested as the closing and reopening of a gap at the Z point. The calculated surface spectra of Fermi surface (Figs.~\refer{StrainCalc}{f-h}) and energy dispersions (Figs.~\refer{StrainCalc}{k-p}) across \X{} and \Gama{} both reproduce well our observations of two-step phase transitions from a strong TI phase to a trivial semimetal phase, and moreover to a trivial band insulator phase, with increasing the tensile strain up to 2.5 $\%$, which is comparable to the maximum strain value in our experiments; the spectral agreement between experimental observations (Figs.~\refer{Strain}{f,k,q} in semimetal phase and Figs.~\refer{Strain}{g,m,s} in insulator phase) and calculations (Figs.~\refer{StrainCalc}{g,m,n} and Figs.~\refer{StrainCalc}{h,o,p}, respectively) is excellent (see Supplementary Figs.~\zref{SemimetalFig},\zref{InsulatorFig} for direct comparisons). 

We also calculated the effect of compressive strain on \TS{} (Figs.~\refer{StrainCalc}{a,e,i,j}); 
it not only enlarges the band inversion at the Z point, but also inverts another band at the B point (Fig.~\refer{StrainCalc}{a}), triggering an additional topological phase transition from a strong TI phase to a weak TI phase, where the TSSs emerge only at the side surface (Fig.~\refer{StrainCalc}{q}). Applying compressive strain to quasi-1D crystals is experimentally very challenging since a compression in the chain direction easily bends or wrinkles the crystals, which releases the compressive force before reaching the amount we intended to apply. Even so, we have applied compressive strain to some extent and successfully observed the enhancement of band inversion, which indicates that the system was driven towards the weak TI phase by the compressive strain, as calculations predict (see Supplementary Figs.~\zref{CompressiveStrainFig} and \zref{CompressiveWTIFig}); nevertheless, the decisive evidence for the weak TI phase is lacking, so we leave it for a future work.

In Fig.~\refer{StrainCalc}{r}, we summarize the evolution of the direct energy gaps at B and Z, and 
also the indirect gap of the system, with strain applied along the chain $\delta_{b}$($\%$), which determine the bulk topology and the metallic nature of \TS{}, respectively. The topological quantum critical points have been obtained at $\delta_{b}$ \wave{} -0.75 \% and \wave{} 0.95 \%; the band inversion occurs both at B and Z for $\delta_{b} <$ -0.75 \%, and only at Z for -0.75 \% $< \delta_{b} <$ 0.95 \%, and eventually it is ruled out for $\delta_{b} >$ 0.95 \%, where a weak TI, a strong TI, and a trivial semimetal phase appear, respectively. Moreover, we also found another type of critical point  at $\delta_{b}$ \wave{} 1.75 \%, across which the metal-insulator transition occurs. The correspondence between the $\delta_{b}$ values applied and the electronic phases obtained in calculations matches surprisingly well with our experimental results (Fig.~\refer{Strain}{v}). 

The strain-induced topological phase transition can be generally interpreted as follows. The tensile strain increases the distance among atoms building the crystal, which suppresses the hopping integral of electrons and thus shrinks the electronic bands. As a result, the overlapping of conduction and valence bands is diminished, and eventually, the band inversion is lifted with enough strain, causing the transition from the non-trivial topological phase to the trivial phase. The calculations in Figs.~\refer{StrainCalc}{b-d} clearly show that the valence bands are narrowed as tensile strain increases, lifting the band inversion and eventually opening the full band gap. The opposite occurs with the compressive strain (Fig.~\refer{StrainCalc}{a}): the electronic bands expand due to the larger hopping integral of electrons, which enhances the overlapping of conduction and valence bands. Consequently, another band inversion occurs, causing the transition from the strong TI phase to the weak TI phase.

The quasi-1D topological state of a superconductor \TS{} we have revealed could have several different scientific and technological implications. In particular, \TS{} has an excellent functionality in that the band topology is easily controlled by mechanical strain, leading to the reversible on/off switching of highly directional, dense spin currents that are protected against the impurity scattering (see Supplementary Fig.~\zref{StrainControlFig} for the reversibility). To date, the topological phase transition from a strong TI phase to a trivial insulator phase has been observed directly by ARPES for alloys, where the bulk topology is controlled by elemental substitution, which varies both the lattice constant and the SOC strength. A new way for it we have demonstrated here, which tunes only the lattice constant by strain, is much simpler, reversible, and thus promising for the future application. Hence, our experimental observations of the topological phase transition between a strong TI state and a trivial state realized in a superconductor will prompt further basic and technological researches for engineering the topological spintronics or optoelectronics which exploits the topological phase transition and the interplay of topological state with superconductivity likely leading to the topological superconducting state. While further works are required, the expectation for the availability to control the Majorana fermions would encourage follow-up researches on the new strategies for realizing quantum computation.

\listoffiguresbookmarks

\bigskip

\pdfbookmark[1]{Methods}{Methods}
\textbf{Methods}
\bigskip

\pdfbookmark[2]{Samples}{Samples}
\textbf{Samples}

High-quality monoclinic whisker crystals of \TS{} were synthesized by the vapor phase transport method, with a = 10.405 \A{}, b = 3.499 \A{}, c = 9.826 \A{}, and $\beta$ = 106.26\si{\degree}~\cite{Nomura_EPL1191_2017}.
Typical samples used for the ARPES and SARPES measurements are shown in Fig.~\refer{ARPES}{c} with in-plane dimension \wave{} 500 $\times$ 50 \textendash~100 \si{\square\um}.
The samples used in the strain-dependent measurements were intentionally selected to be as small and thin as possible (generally 500 \si{\times} 50 \si{\times} 2 \si{\cubic\um}), to diminish any strain relaxation effects.
More details can be found in Supplementary Note~\zref{Samples}.\\

\pdfbookmark[2]{ARPES}{ARPES}
\textbf{ARPES}

Synchrotron-based ARPES measurements were performed at the beamline 5-2 of Stanford Synchrotron Radiation Lightsource (SSRL) with tunable photon energies from 25 eV to 100 eV.
The spot size was \wave{} 0.04 (horizontal) $\times$ 0.01 (vertical) \si{\square\mm} and the equipped ScientaOmicron DA30L analyser enables a full access of the momentum space without sample rotation, both facilitating the measurements as the typical sample width was \wave{} 50 \textendash~100 \si{\um}.
The light polarization used was linear vertical (along the analyser slit) with the overall energy resolution from 10 \textendash~20 meV and angular resolution better than 0.2\si{\degree}; see Supplementary Notes~\zref{hvGamma},\zref{hvX} and Figs.~\zref{hvGammaFig},\zref{hvXFig} for more detailed discussions about photon energy and light polarization dependences of the spectra.
The samples were cleaved using exfoliation method and measured \IS{} at a temperature \wave{} 10 K and a pressure better than \num{3d-11} Torr.

The laser-based SARPES~\cite{Yaji_Rev.Sci.Instrum.875_2016} measurements were performed at the Institute for Solid State Physics, University of Tokyo, equipped with 6.994 eV photons and a ScientaOmicron DA30L analyser.
The laser was $p$-polarized with spot size \wave{} 0.05 \si{\square\mm}.
See Supplementary Notes~\zref{SpinComponent},\zref{SpinLaserPolarization} and Figs.~\zref{SpinComponentFig},\zref{SpinLaserPolarizationFig} for more information about the laser polarizations and apparatus geometry.
The energy (angular) resolution was set at better than 2 meV (0.2 \si{\degree}) and 15 meV (0.7\si{\degree}) for ARPES and SARPES, respectively.
The samples were cleaved using similar method at room temperature with a pressure better than \num{5d-9} Torr and measured at \wave{} 10 K with a pressure better than \num{7d-11} Torr.
The spin-polarization map in Fig.~\refer{Laser}{l} utilizes a 2D color coding with horizontal and vertical axes indicating the spectral weight and the polarization, respectively~\cite{Tusche_Ultramicroscopy159_2015,Noguchi_Phys.Rev.B954_2017}.

The strain-dependent laser-ARPES measurements were performed in similar conditions to SARPES but with a ScientaOmicron R4000 analyser.
The angular resolution was \wave{} 0.3\si{\degree} and the laser polarization was linear horizontal (along the analyser slit).
More information about laser polarizations and apparatus geometry is shown in Supplementary Fig.~\zref{hvGammaFig}.
The samples were mounted on the strain device using silver epoxy (Muromac H-220), cleaved \IS{} at room temperature with a pressure better than \num{1d-10} Torr, and measured at a temperature $<$ 10 K with a pressure \wave{} \num{1d-11} Torr.
The \IS{} strain controls were realized by rotating the screws 45\si{\degree} step by step at a temperature $<$ 30 K with a pressure better than \num{5d-10} Torr.
Typical time for one 45\si{\degree} rotation was \wave{} 2 minutes and we confirmed no surface aging by performing strain control cycles and measuring at freshly cleaved samples with strain applied \textit{ex-situ} (Supplementary Fig.~\zref{StrainControlFig}).
We also show the \IS{} strain-control reversibility test in Supplementary Fig.~\zref{StrainControlFig}, the results indicate the strain effects are completely reversible.
After each measurement, we monitored the device status, estimated and calibrated the \IS{} controlled strain magnitude to the strain measured by the strain gauges, as described below.
\bigskip

\pdfbookmark[2]{Strain evaluation}{Strain evaluation}
\textbf{Strain evaluation}

The strain device~\cite{Ricco_Nat.Commun.91_2018} was specially designed for the low-energy laser-ARPES apparatus.
The key components of the device include two BeCu bridges, one BeCu sample platform (10 \si{\times} 2 \si{\times} 1 \si{\cubic\mm}), and one $\text{Si}_3\text{N}_4$ ball (1 \si{\mm} diameter) below it.
When the screws are tightened, the sample platform is bended by the bridges providing uniaxial strain to the samples mounted, as demonstrated in Fig.~\refer{Strain}{b}.
As the laser spot size is small, we found no evidence of spectral broadening due to possible domains introduced by bending.
Good thermal (electric) conductivity has been confirmed by comparing the spectra with those obtained from samples mounted on normal sample holders.

The strain magnitude as a function of screw angles (Fig.~\refer{Strain}{d}) was measured by the commercial strain gauges (Fig.~\refer{Strain}{c}, KYOWA KFGS-1N-120-C1-11) at room temperature.
For both the strain gauge measurements and the strain-dependent ARPES measurements, we always carefully adjusted the screws and platforms to ensure the initial conditions of the devices were identical within error bars: before the measurements, the heights of the fours screws and the bridges were equally adjusted to ensure a flat platform;
meanwhile, the displacements between the center of the sample platform and its contact positions with the bridges were carefully monitored, so that a zero net strain was ensured before attaching the strain gauges or mounting samples.
We have conducted several strain gauge measurements on different devices and obtained similar results (two of them are shown in Fig.~\refer{Strain}{d}).
By fitting the data with a polynomial function (the red curve in Fig.~\refer{Strain}{d}), we can then estimate the strain magnitude \IS{} controlled during the ARPES measurements by interpolating at the corresponding screw angles.
We also measured the displacements as a function of the strain, as shown in Supplementary Fig.~\zref{StrainDeviceFig}d.
Comparing to the large strain in our experiments (\wave{} 2.4\%), we ignore the thermal contraction which is typically \wave{} -0.2\% according to our thermal-strain simulations. 
The error bars for the strain magnitude controlled in the ARPES measurements were estimated in accordance with the uncertainty of screw angles and shown in Figs.~\refer{Strain}{d,v} (smaller than the marker size in Fig.~\refer{Strain}{d}).

The strain simulations based on finite element analysis were performed when the platform displacement reached 0.124 \si{\mm} (Supplementary Figs.~\zref{StrainDeviceFig}d,g), using the Autodesk Inventor 2020.
For the BeCu components, a Young's modulus of 120 GPa and a Poisson's ratio of 0.3 were implemented.
Figure~\refer{Strain}{b} and Supplementary Fig.~\zref{StrainDeviceFig}e show the simulated axial strain ($\epsilon_\text{xx}$) distribution on the sample platform.
$\epsilon_\text{xx}$ reached the maximum \wave{} 2.4\% at the platform center, nicely consistent with our experimental results from the strain gauges at the same displacement (Supplementary Fig.~\zref{StrainDeviceFig}d).
The transverse strain ($\epsilon_\text{yy}$) distribution is shown in Supplementary Fig.~\zref{StrainDeviceFig}f, which reached the minimum \wave{} -0.2\% at the platform center.
Supplementary Fig.~\zref{StrainDeviceFig}h summarizes the overall strain distribution by plotting the $\epsilon_\text{xx}$ and $\epsilon_\text{yy}$ as a function of coordinates along x (y = 0) and along y (x = 0), respectively.
We note that the samples used for strain-dependent ARPES measurements were very small (generally 500 \si{\times} 50 \si{\times} 2 \si{\cubic\um}) and mounted at the center of the platform.
Therefore, the tiny transverse strain perpendicular to the platform is neglectable.
Moreover, as the samples used were also very thin, our simulations show that the strain on the sample surface was almost the same as that on the platform center (Supplementary Figs.~\zref{StrainDeviceFig}i,j).
See Supplementary Notes~\zref{StrainDevice},\zref{StrainControl} for more detailed descriptions about the strain devices, strain evaluation, and \IS{} strain controls.

\bigskip

\pdfbookmark[2]{Calculations}{Calculations}
\textbf{DFT Calculations}

First, we performed the structural optimization using the Perdew-Burke-Ernzerhof exchange-correlation functional revised for solids~\cite{Perdew_Phys.Rev.Lett.10013_2008} and the projector augmented wave method~\cite{Kresse_Phys.Rev.B593_1999} as implemented in {\it Vienna ab initio Simulation Package}~\cite{Kresse_Phys.Rev.B471_1993,Kresse_Phys.Rev.B4920_1994,Kresse_Comput.Mater.Sci.61_1996,Kresse_Phys.Rev.B5416_1996}.
The spin-orbit coupling was included in all the calculations presented in this paper.
Here, we optimized atomic coordinates while the lattice parameters were fixed as our experimental values.
We also optimized atomic coordinates in strain calculations, where the lattice parameters except for $b$ (the length along the strain direction) were fixed as unstrained values for simplicity.
Next, we performed first-principles band-structure calculation using the modified Becke-Johnson potential~\cite{Tran_Phys.Rev.Lett.10222_2009} as implemented in the \textsc{WIEN2k} code~\cite{Blaha__2018}.
The muffin-tin radii $r$ for Ta and Se were set to 2.5 and 2.39 a.u., respectively, for the unstrained calculation while they are slightly changed for the strain calculations.
The maximum modulus for the reciprocal lattice vectors $K_{\mathrm{max}}$ was chosen so that $r_\text{Se}K_{\mathrm{max}}=8$.
We took a 4$\times$12$\times$4 $\bm{k}$-mesh for a self-consistent-field (SCF) calculation, while a following non-SCF calculation to determine the Fermi energy was performed using a finer $\bm{k}$-mesh up to 25,000 points.
From the calculated band structures, we extracted the Wannier functions of the Ta-$s,d$ and Se-$s,p$ orbitals using the \textsc{Wien2Wannier} and \textsc{Wannier90} codes~\cite{Marzari_Phys.Rev.B5620_1997,Souza_Phys.Rev.B653_2001,Kunes_ComputerPhysicsCommunications18111_2010,Pizzi_J.Phys.:Condens.Matter3216_2020}.
We did not perform the maximal localization procedure for the Wannier functions to prevent orbital mixing among the different spin components.
The semi-infinite-slab tight-binding models constructed from these Wannier functions were used for calculating the surface spectra in the way described in Ref.~\onlinecite{Sancho_J.Phys.F:Met.Phys.154_1985}.
The Fermi energy in all the calculated data is shifted by -10 meV so as to reproduce the experimental spectra.

The tensile strain along the chain direction leads to finite compressive strain perpendicular to it due to the Poisson effect. However, we did not take the perpendicular compressive strain into account in the DFT calculations of Fig.~\ref{StrainCalc} since the effect is negligible and not crucial for our conclusions. To confirm this, we have performed a simulation considering the Poisson’s ratio and estimated the perpendicular strain to be only -0.2\% when the tensile strain along the chain direction was set to 2.4\% (Supplementary Fig.~\zref{StrainDeviceFig}h), which is the maximum value we applied for the samples in our experiments. Subsequently, we calculated the bulk band structure under the 2.4\% tensile strain along the chain direction with and without the effect of the perpendicular compressive strain of -0.2\%; no considerable difference was obtained between the two (Supplementary Fig.~\zref{BulkCalcPoissonFig}), leading us to conclude that the effect of the perpendicular compressive stain can be ignored in our study.

\bigskip

\textbf{Data availability}

The data that support the findings of this study are available from the corresponding author upon request.

\bigskip

\textbf{Acknowledgements}

We thank D. Hirai and Y. Mizukami for fruitful comments on the strain measurements, and also thank S. Sakuragi and T. Yajima for XRD measurements. Use of the Stanford Synchrotron Radiation Lightsource, SLAC National Accelerator Laboratory, is supported by the U.S. Department of Energy, Office of Science, Office of Basic Energy Sciences under Contract No. DE-AC02-76SF00515.
This work was supported by the JSPS KAKENHI (grant numbers JP18H01165, JP18K03484, JP19H02683, JP19F19030 and JP19H00651), MEXT Q-LEAP (grant number JPMXS0118068681), and by MEXT under the “Program for Promoting Researches on the Supercomputer Fugaku” (Basic Science for Emergence and Functionality in Quantum Matter Innovative Strongly Correlated Electron Science by Integration of “Fugaku” and Frontier Experiments) (Project ID: hp200132).

\bigskip

\textbf{Author contributions}

T.K. and S.T. planned the experimental project.
C.L. conducted ARPES experiments, analysed the data, and performed strain simulations. 
C.L. and T.K. designed the strain devices.
C.L. Ke.K, Y.A., and T.K. conducted strain gauge measurements.
R.N., Ke.K., P.Z., C.B., Ki.K., Y.A., Ka.K., H.T., K.Y., A.H., M.H., D.L., S.S., and T.K. supported ARPES experiment.
M.S., A.N., M.T., and S.T. prepared the single crystals.
M.O. and R.A. calculated and analyzed the theoretical band structure.
C.L., M.O., and T.K. wrote the paper.
All authors discussed the results and commented on the manuscript.

\bigskip

\textbf{Competing interests}

The authors declare no competing interests.

\bigskip

\textbf{Additional information}

Correspondence and requests for materials should be addressed to T.K. (email: kondo1215@issp.u-tokyo.ac.jp).

%

\end{document}